\documentclass[pra,twocolumn]{revtex4}

\usepackage{graphics}
\usepackage{graphicx}
\usepackage{bm}
\usepackage{latexsym}
\usepackage{color}
\usepackage{mathrsfs}
\usepackage{braket}
\usepackage{enumerate}
\usepackage[usenames,dvipsnames]{xcolor}
\usepackage[colorlinks=true,citecolor=blue,linkcolor=blue,urlcolor=blue,backref=true]{hyperref}
\usepackage{float}
\usepackage[raggedright,tight]{subfigure}  
\usepackage{amsmath}
\usepackage{amsfonts}
\usepackage{braket}
\usepackage[dvipsnames]{xcolor}
\usepackage{rotating}
\usepackage{multirow}
\usepackage{xcolor}

\begin{document}

\title{Identification of Poincar\'e-gauge and multipolar nonrelativistic theories of QED}
\author{Adam Stokes}\email{adamstokes8@gmail.com}
\author{Ahsan Nazir}
\affiliation{Department of Physics and Astronomy, University of Manchester, Oxford Road, Manchester M13 9PL, United Kingdom}

\date{\today}

\begin{abstract}
{For over six decades, quantum electrodynamics (QED) in multipolar form has been an invaluable tool for understanding quantum-scale atomic and molecular interactions. However, its relation to the Poincar\'e-gauge has been a recent topic of controversy and debate. It was claimed by Rousseau and Felbacq in the article Scientific Reports \textbf{7}, 11115 (2017) that Hamiltonian multipolar QED is not the same as Poincar\'e-gauge QED and that it is not generally equivalent to Coulomb-gauge QED. This claim has subsequently been refuted, but since both sides of the debate appear technically sound, a clear reconciliation remains to be given. This task is of paramount importance due to the widespread use of multipolar QED in quantum optics and atomic physics. Here, unlike in other responses, we adopt the same method as Rousseau and Felbacq of using Dirac's constrained quantisation procedure. However, our treatment shows that Poincar\'e-gauge and multipolar QED are identical. We identify the precise source of the apparent incompatibility of previous results as nothing more than a semantic mismatch. In fact there are no inconsistencies. Our results firmly and rigorously solidify the multipolar theory.} 
\end{abstract}

\maketitle


\section{Introduction}

Quantum electrodynamics (QED) is the prototypical quantum gauge-field theory. Therein, the non-dynamical constraint called Gauss' law; $\nabla\cdot {\bf E}=\rho$ where ${\bf E}$ is the electric field and $\rho$ the density of charges, implies gauge redundancy. When electromagnetic potentials are used as generalised coordinates the Lagrangian of the theory is degenerate implying that there will be fewer canonical momenta than canonical coordinates when passing to the Hamiltonian formalism. Dirac laid out a systematic procedure by which a suitable Lie algebra of classical observables, known as Dirac brackets, can be constructed, which respect both the equations of motion and the non-dynamical constraints. This enables passage to the quantum theory via the replacement of Dirac brackets with commutators. Applied to QED, Dirac's procedure requires invoking a gauge-fixing constraint to eliminate gauge redundancy, which identifies a physical subspace of states. 

In the non-relativistic theory of atoms and molecules QED in multipolar form has been an invaluable tool over the past six decades and has been found to agree well with experiment. Indeed, the first example of this goes back over fifty years to the pioneering work of Power and Zienau \cite{power_coulomb_1959} in predicting the natural lineshape of atomic hydrogen within Lamb's famous experiments \cite{lamb_fine_1952}. In the vast recent literature, multipolar QED has seen numerous applications, for example, \cite{woolley_molecular_1971,woolley_reformulation_1974,woolley_non-relativistic_1975,Javanainen_exact_2017,quesada_why_2017,stokes_implications_2021,stokes_ultrastrong_2021,stokes_gauge_2019,drezet_dual-lagrangian_2016,grinter_resonance_2016,messina_dynamical_2010,stokes_gauge_2013,stokes_noncovariant_2012,komninos_theory_2017,forbes_optical_2018,chernyak_non-linear_2015,tokman_purcell_2019,de_bernardis_breakdown_2018,stokes_extending_2012,safari_van_2008,horsley_role_2008,brooke_super-_2008,cho_single_2008,pipolo_cavity_2014}. As photonic systems continue to diversify and the field continues to expand, multipolar QED will continue to serve as a fundamentally important theoretical tool.

The derivation of multipolar QED and its relation to Coulomb-gauge QED is standard textbook material \cite{cohen-tannoudji_photons_1989,craig_molecular_1998}. Nearly always however, the multipolar form is obtained as a Power-Zienau-Woolley (PZW) transformation of the Coulomb-gauge theory, either at  the Lagrangian or Hamiltonian level. In other words, the Coulomb-gauge is fixed first to obtain the physical subspace and one then works from there. Automatically, this method means that the multipolar theory is expressed in terms of the transverse vector potential, which is a  gauge-invariant quantity, but which is often mistaken as belonging to the Coulomb-gauge because of the property that the longitudinal vector potential vanishes therein. The PZW transformation also transforms the Coulomb-gauge canonical momentum operators, which therefore possess different physical meaning before and after the transformation. In the Coulomb-gauge the field canonical momentum is the transverse electric field. In the multipolar formalism however, the momentum conjugate to the transverse vector potential is not the transverse electric field. 

Multipolar QED is clearly closely related to the so-called Poincar\'e gauge, but due to the preceding facts, some confusion has arisen regarding whether and in what sense multipolar QED can really be considered an alternative gauge choice to the Coulomb-gauge. Naively speaking, a different gauge choice to the Coulomb-gauge would presumably entail a theory written in terms of a different vector potential. Moreover, it is not immediately obvious why the field canonical momentum should change from the transverse electric field. Yet multipolar QED seems to contradict these naive expectations. Indeed, Refs.~\cite{rousseau_quantum-optics_2017,rousseau_reply_2018} apply Dirac's method in the case of nonrelativistic material charges to derive the Poincar\'e-gauge canonical theory and they claim that textbook multipolar QED is in fact not the same as the correct Poincar\'e-gauge theory. They claim further that the multipolar theory will not produce the same results as the well-known Coulomb-gauge theory. Refs.~\cite{vukics_gauge-invariant_2021} and \cite{andrews_perspective_2018} dispute this claim, concluding that criticisms of the multipolar framework in Ref.~\cite{rousseau_quantum-optics_2017} are not valid. In turn Ref.~\cite{rousseau_reply_2018} disputes the conclusions of Ref.~\cite{vukics_gauge-invariant_2021} maintaining that the conclusion of Ref.~\cite{rousseau_quantum-optics_2017} is valid. 

Although the Refs. \cite{vukics_gauge-invariant_2021} and \cite{andrews_perspective_2018} contain valuable insights, the situation has not been clearly and decisively resolved. This is because it is not clear where, if anywhere, either side of the debate is technically flawed, and if both sides are technically sound, it is unclear how the apparent disagreement comes to be. Here, we adopt the same method of Dirac's constrained quantisation as is used in Refs.~\cite{rousseau_quantum-optics_2017,rousseau_reply_2018}. 
We show precisely why this claim is in fact not valid, despite the technical validity of the Poincar\'e-gauge theory obtained in  Refs.~\cite{rousseau_quantum-optics_2017,rousseau_reply_2018}. We demonstrate how the theory derived in this way is in fact identical to textbook multipolar theory. The key to the resolution we provide is the construction of the canonical operators that are commonly used in multipolar QED {\em from} the canonical operators of the Poincar\'e-gauge that are found using Dirac's method. Attempting to equate these distinct canonical operator sets results in the incorrect conclusion that the two theories are disparate. In fact the two sets of operators are not identical, but they can be easily related. Expressing the Poincar\'e-gauge theory in terms of the multipolar canonical operators reveals that the two theories are identical. We show moreover, that all fixed-gauge theories of QED are equivalent by explicitly constructing the necessary unitary gauge-fixing transformations that act within the physical state space. The Power-Zienau-Woolley transformation between Coulomb-gauge and multipolar QED is an example of such a transformation.

This article is divided into four sections. In Sec.~\ref{back} we provide a pedagogical overview of gauge-freedom, which facilitates a straightforward understanding of our results in Sec.~\ref{results}. This includes understanding both electromagnetic and material auxiliary potentials. In Sec.~\ref{results} we provide our main results, which resolve all controversy surrounding multipolar QED and its relation to the Poincar\'e-gauge, rigorously solidifying the multipolar theory. In Sec.~\ref{discuss} we generalise our results by providing the Hamiltonian in an arbitrary gauge. We show that all such Hamiltonians are unitarily equivalent and provide the general form of unitary gauge-fixing transformations of which well-known Power-Zienau-Woolley (PZW) transformation is an example. We summarise our findings briefly in Sec.~\ref{conc}.

\section{Background}\label{back}

\subsection{Gauge-freedom and electromagnetic potentials}

Throughout this article we use natural Lorentz-Heaviside units. For simplicity we restrict our attention to a single-electron atom with fixed nucleus at the origin within the surrounding electromagnetic field. The classical charge and current densities are
\begin{align}
&\rho({\bf x})=q\delta({\bf x}-{\bf r})-q\delta({\bf x})\label{rho}\\
&{\bf J}({\
\bf x})=q{\dot {\bf r}}\delta({\bf x}-{\bf r})
\end{align}
where $q$ is the electron's charge, and ${\bf r}$ is its position. Electric and magnetic fields ${\bf E}$ and ${\bf B}$ are defined in terms of the scalar and vector potentials $A_0$ and ${\bf A}$ as
 \begin{align}
{\bf E}&=-\nabla A_0 - {\dot {\bf A}}\label{elecf},\\
{\bf B}&=\nabla\times {\bf A}\label{magf}.
\end{align}
These definitions imply that the homogeneous Maxwell equations, $\nabla \cdot {\bf B}=0$ and ${\dot {\bf B}} = -\nabla\times {\bf E}$, are automatically satisfied. To see this note that $\nabla\cdot \nabla\times {\bf V}=0$ for any twice differentiable ${\bf V}$ and that $\nabla \times \nabla V={\bf 0}$ for any twice differentiable $V$. The inhomogeneous constraint $C_1:= \nabla \cdot {\bf E}-\rho=0$ (Gauss' law) must be imposed within the theory while the remaining inhomogeneous equation is dynamical ${\dot {\bf E}}=\nabla \times {\bf B}-{\bf J}$ (Maxwell-Ampere law). This is an equation of motion that must be produced by any satisfactory Lagrangian or Hamiltonian description. 

The electric and magnetic fields are invariant under the {\em gauge transformation}
\begin{align}
{\bf A}'&={\bf A}+\nabla\chi,\label{gt1}\\
A_0'&=A_0-{\dot \chi}
\end{align}
where $\chi$ is an arbitrary function over spacetime. Recall that the Helmholtz decomposition of a vector-field ${\bf V}$ into transverse and longitudinal fields, ${\bf V}={\bf V}_{\rm T}+{\bf V}_{\rm L}$, is unique. The transverse and longitudinal components satisfy $\nabla\cdot {\bf V}_{\rm T}=0$ and $\nabla\times {\bf V}_{\rm L}={\bf 0}$. We see therefore that the transverse vector potential ${\bf A}_{\rm T}$ is gauge-invariant and unique, that is, if ${\bf A}$ and ${\bf A}'$ are related as in Eq.~(\ref{gt1}) then ${\bf A}_{\rm T}'={\bf A}_{\rm T}$. Gauge-freedom is therefore the freedom to choose the longitudinal vector potential ${\bf A}_{\rm L}=\nabla\chi$ where ${\bf A}={\bf A}_{\rm T}+\nabla\chi$. Perhaps the most straightforward choice is ${\bf A}_{\rm L}={\bf 0}$, which is called the Coulomb-gauge. If we denote the corresponding Coulomb-gauge scalar potential by $\phi$ then we have from Eq.~(\ref{elecf}) that ${\bf E}=-\nabla \phi-{\dot {\bf A}}_{\rm T}$ from which it follows that $\nabla \cdot {\bf E} = -\nabla^2 \phi$. If we invoke Gauss' law, $\nabla\cdot {\bf E}=\rho$, then we obtain
\begin{align}
\phi({\bf x}) = -{1\over \nabla^2}\rho({\bf x}) = \int d^3 x' {\rho({\bf x}')\over 4\pi|{\bf x}-{\bf x}'|},
\end{align} 
which is called the Coulomb potential of the charge distribution $\rho$.

The Coulomb-gauge potentials $(\phi,{\bf A}_{\rm T})$ provide a convenient reference set in terms of which any other gauge may be specified as
\begin{align}
&{\bf A}={\bf A}_{\rm T}+\nabla \chi,\label{at}\\
&A_0 = \phi-{\dot \chi}.\label{phit}
\end{align}
We emphasise however, that it is incorrect to identify ${\bf A}_{\rm T}$ and $\phi$ as {\em belonging} to the Coulomb-gauge, because they are well-defined fields that are identifiable and the same in every gauge. What defines the Coulomb-gauge is the (gauge-fixing) condition that $\chi$ in Eqs.~(\ref{at}) and (\ref{phit}) vanishes. This condition has nothing to do with $\phi$ and ${\bf A}_{\rm T}$. Its effect is to fix ${\bf A}$ as equal to ${\bf A}_{\rm T}$ and to fix $A_0$ as equal to $\phi$, but whether or not these equalities happen to hold, the quantities ${\bf A}_{\rm T}$ and $\phi$ are always well-defined and identifiable. 

Since the transverse potential ${\bf A}_{\rm T}$ is gauge-invariant, in any gauge it can be used as an elementary physical coordinate for the electromagnetic field. The standard Coulomb and Poincar\'e gauges of nonrelativistic QED can be generalised by specifying the freely choosable gauge-function $\chi$ as a functional of ${\bf A}_{\rm T}$. 
This can be achieved by defining the gauge-fixing constraint
\cite{woolley_r._g._charged_1999}
\begin{align}\label{c2}
C_2:=\int d^3 x' {\bf g}({\bf x}',{\bf x})\cdot {\bf A}({\bf x'}) =0
\end{align}
in which ${\bf g}$ is the Green's function for the divergence operator;
\begin{align}
\nabla\cdot {\bf g}({\bf x},{\bf x}')=\delta({\bf x}-{\bf x}').
\end{align}
The longitudinal part ${\bf g}_{\rm L}({\bf x},{\bf x}')$ is uniquely defined by this equation as the gradient of the Green's function for the Laplacian [cf. Eq.~\ref{gl}];
\begin{align}\label{gl}
{\bf g}_{\rm L}({\bf x},{\bf x}')= -\nabla{1\over 4\pi|{\bf x}-{\bf x}'|}
\end{align}
whereas the transverse part ${\bf g}_{\rm T}({\bf x},{\bf x}')$ is a completely arbitrary function of ${\bf x}'$ and an arbitrary transverse function of ${\bf x}$. Given the constraint $C_2=0$, choosing a concrete ${\bf g}_{\rm T}$ specifies the gauge, because imposing $C_2=0$ implies that ${\bf A}$ can be written \cite{woolley_r._g._charged_1999}
\begin{align}\label{Aarb}
{\bf A}({\bf x})={\bf A}_{\rm T}({\bf x})+\nabla\int d^3 x' {\bf g}({\bf x'},{\bf x})\cdot {\bf A}_{\rm T}({\bf x}'),
\end{align}
which defines the gauge-function $\chi_g$ such that $\nabla\chi_g={\bf A}_{\rm L}$ as
\begin{align}
\chi_g({\bf x}) =& \int d^3 x' {\bf g}({\bf x'},{\bf x})\cdot {\bf A}_{\rm T}({\bf x}') \nonumber \\=& \int d^3 x' {\bf g}_{\rm T}({\bf x'},{\bf x})\cdot {\bf A}_{\rm T}({\bf x}').
\end{align}
The Coulomb-gauge is defined by ${\bf g}_{\rm T}={\bf 0}$. The Poincar\'e-gauge is defined by the condition ${\bf x}\cdot {\bf A}({\bf x})=0$ and it is easily verified that by letting
\begin{align}\label{pgt}
{\bf g}_{\rm T}({\bf x},{\bf x}') = -\int_0^1 d\lambda\, {\bf x}' \cdot \delta^{\rm T}({\bf x}-\lambda{\bf x}')
\end{align}
in Eq.~(\ref{Aarb}) we obtain a potential for which ${\bf x}\cdot {\bf A}({\bf x})=0$. Here the transverse $\delta$-function $\delta^{\rm T}$ is defined by
\begin{align}
\delta_{ij}^{\rm T}({\bf x}) = \delta_{ij}\delta({\bf x})-\delta_{ij}^{\rm L}({\bf x}) 
\end{align}
where
\begin{align}\label{dlong}
\delta_{ij}^{\rm L}({\bf x}) = -\nabla_i \nabla_j {1\over 4\pi|{\bf x}|} = \int {d^3k\over (2\pi)^3} \, {\hat k}_i{\hat k}_j e^{i{\bf k}\cdot {\bf x}}
\end{align}
defines the longitudinal $\delta$-function $\delta^{\rm L}$.

Although the gauge-fixing constraint $C_2=0$ in Eq.~(\ref{c2}) is certainly not sufficiently general to include all possible gauge choices, as shown above, it does include both the Coulomb and Poincar\'e gauges as special cases. It is therefore suitable for our purpose of clarifying the relationship between the Coulomb-gauge, the Poincar\'e-gauge, and the multipolar theories of non-relativistic QED. In Secs.~\ref{results} and \ref{discuss} we will use Dirac's constrained quantisation procedure in conjunction with the constraints $C_1=0$ and $C_2=0$, to obtain an arbitrary-gauge Hamiltonian QED, in which the gauge choice is controlled through the choice of the arbitrary function ${\bf g}_{\rm T}$.


\subsection{Gauge-freedom and material potentials}

Before providing our main results we briefly discuss the lesser known gauge-freedom that is inherent in material auxiliary potentials. Doing so already allows us to identify the connection between the Poincar\'e-gauge and the well-known multipolar formalism. Auxiliary material potentials ${\bf P}$ and ${\bf M}$ can be defined using the inhomogenous Maxwell equations;
\begin{align}
&\rho = -\nabla\cdot {\bf P}\label{P},\\
&{\bf J}={\dot {\bf P}}+\nabla \times {\bf M}.
\end{align}
Unlike the electric and magnetic fields ${\bf E}$ and ${\bf B}$ which are defined by the inhomogeneous Maxwell equations and accompanying homogeneous Maxwell equations, ${\bf P}$ and ${\bf M}$ are auxiliary material potentials that are not required to satisfy homogeneous Maxwell equations, and so they are not uniquely specified. They can be viewed as a material analog of the non-unique auxiliary potentials $A_0$ and ${\bf A}$ for the electromagnetic field. Specifically, in the same way that ${\bf E}$ and ${\bf B}$ are invariant under a gauge transformation of $(A_0,{\bf A})$, the physical charge and current densities $\rho$ and ${\bf J}$ are invariant under a transformation of $({\bf P},{\bf M})$ by pseudo-magnetic and pseudo-electric fields as
\begin{align}\label{pmt}
&{\bf P}\to {\bf P}+\nabla \times {\bf U},\\
&{\bf M} \to {\bf M}-\nabla U_0 -{\dot {\bf U}}
\end{align}
where $(U_\mu)=(U_0,-{\bf U})$ are the components of an arbitrary pseudo-four-potential. The polarisation ${\bf P}$ and magnetisation ${\bf M}$ are in turn invariant under a gauge transformation $U_\mu\to U_\mu -\partial_\mu \chi$ where $\mu=0,\,1,\,2,\,3$ and $\chi$ is arbitrary.

The field ${\bf M}_{\rm L}$ is completely arbitrary because it does not contribute to either $\rho$ or ${\bf J}$. Only the transverse freedom in ${\bf P}$ and ${\bf M}$ is non-trivial. By specifying ${\bf P}_{\rm T}$ both ${\bf P}$ and $\nabla\times {\bf M} = {\bf J}_{\rm T}-{\dot {\bf P}}_{\rm T}={\bf J}-{\dot {\bf P}}$ are fully specified. If we define the polarisation ${\bf P}$ as
\begin{align}\label{P2}
{\bf P}({\bf x}) = -\int d^3 x' {\bf g}({\bf x},{\bf x}')\rho({\bf x}') 
\end{align}
then we see that $-\nabla \cdot {\bf P} = \rho$ is satisfied identically and ${\bf P}_{\rm L}=\nabla \phi$ (the gradient of the Coulomb-potential) is obtained from the expression for ${\bf g}_{\rm L}$ given in Eq.~(\ref{gl}). Using Eq.~(\ref{dlong}), it is easy to show via the gradient theorem or by Fourier transformation, that for $\rho$ specified by Eq.~(\ref{rho}), ${\bf P}_{\rm L}$ can also be written as a line-integral between the two charges as
\begin{align}\label{PL}
{\bf P}_{\rm L}({\bf x}) = \int_0^1 d\lambda \, q{\bf r}\cdot \delta^{\rm L}({\bf x}-\lambda{\bf r}).
\end{align}

The transverse polarisation ${\bf P}_{\rm T}={\bf P}-{\bf P}_{\rm L}$ is freely choosable and is fully specified by choosing ${\bf g}_{\rm T}({\bf x},{\bf x}')$, which, under the constraint $C_2=0$, is also what specifies the gauge of the electromagnetic potentials. If, in particular, we choose the Poincar\'e-gauge, then ${\bf g}_{\rm T}$ is given by Eq.~(\ref{pgt}) and upon using Eq.~(\ref{PL}) we see that ${\bf P}$ is nothing but the well-known {\em multipolar} polarisation field for the atom;
\begin{align}\label{multP}
{\bf P}({\bf x}) = \int_0^1 d\lambda \, q{\bf r}\delta({\bf x}-\lambda{\bf r}).
\end{align}
In the following section, our proof that Poincar\'e-gauge and multipolar QED are identical uses the transverse polarisation field ${\bf P}_{\rm T}$ to express the Poincar\'e-gauge theory in terms of the same canonical degrees of freedom that are typically used within textbook expressions of multipolar QED.

\section{Results}\label{results}

We now derive an arbitrary-gauge Hamiltonian quantum theory of the atom-field system via the construction of Dirac brackets. We begin with the standard QED Lagrangian for the nonrelativistic atom within the field \cite{cohen-tannoudji_photons_1989}
\begin{align}\label{L}
L=&L_{\rm KE} + \int d^3 x \,{\cal L}  \nonumber \\ \equiv& {1\over 2}m{\dot {\bf r}}^2-\int d^3 x\, \left[J^\mu({\bf x})A_\mu({\bf x}) +{1\over 4}F_{\mu\nu}({\bf x})F^{\mu\nu}({\bf x})\right], \nonumber \\ =&{1\over 2}m{\dot {\bf r}}^2 -\int d^3 x\, \left[\rho({\bf x})A_0({\bf x}) - {\bf J}({\bf x})\cdot {\bf A}({\bf x})\right] \nonumber \\ &+{1\over 2}\int d^3 x \left[{\bf E}({\bf x})^2-{\bf B}({\bf x})^2\right],
\end{align}
where $m$ is the mass of the dynamical charge $+q$, $(J^\mu)=(\rho,{\bf J})$, $(A_\mu)=(A_0,-{\bf A})$ and $F_{\mu\nu}=\partial_\mu A_\nu - \partial_\nu A_\mu$. Here greek indices take values $0,\,1,\,2,\,3$ and repeated indices are summed. $L_{\rm KE}$ is the kinetic energy of the atomic electron while ${\cal L}$ is the sum of the interaction and pure electromagnetic Lagrangian densities. Using ${\bf r}$ and $A^\mu = (A_0,{\bf A})$ as generalised coordinates, the Lagrangian yields the expected Newton-Lorentz and Maxwell-Ampere dynamical equations. 
The naive canonical momenta ${\bf p}$, ${\tilde {\bm \Pi}}$ and ${\tilde \Pi}_0$ conjugate to ${\bf r}$, ${\bf A}$ and $A_0$ respectively, are obtained from the Lagrangian in the usual way as 
\begin{align}
&{\bf p} = {\partial L\over \partial {\dot {\bf r}}} = m{\dot {\bf r}}+q{\bf A}({\bf r}),\label{pcm}\\ 
&{\tilde {\bm \Pi}} = {\delta L \over \delta{\dot {\bf A}}} = {\dot {\bf A}}+\nabla A_0,\\
&{\tilde \Pi}_0={\delta L \over \delta {\dot A}_0}=0.\label{picm}
\end{align}
Note that the canonical momentum conjugate to ${\bf A}$ is ${\tilde {\bm \Pi}}=-{\bf E}$ in agreement with Refs.~\cite{rousseau_quantum-optics_2017,rousseau_reply_2018}. The condition $C_0:={\tilde \Pi}_0=0$ is our first constraint and is what implies that the Lagrangian $L$ is degenerate. From the equalities ${\partial {\cal L}}/ \partial (\nabla A_0) ={\tilde {\bm \Pi}}$ and $\partial {\cal L}/\partial A_0= -\rho$ we obtain the Euler-Lagrange equation for $A_0$ as
\begin{align}
0=\partial_\mu{\partial {\cal L} \over \partial (\partial_\mu A_0)} - {\partial {\cal L}\over \partial A_0} = {\dot {\tilde \Pi}}_0 +  \nabla \cdot {\tilde {\bm \Pi}} +\rho.
\end{align}
Thus, our second constraint $C_1=0$ (Gauss' law) ensures that if $C_0=0$ at a fixed time, then $C_0=0$ for all times.

The construction of an unconstrained quantum theory proceeds by positing Poisson brackets for the naive Hamiltonian theory as
\begin{align}
\{r_i,p_j\}=&\delta_{ij}, \\ \{A_\mu({\bf x}),{\tilde \Pi}_\nu({\bf x}')\}=&\delta_{\mu\nu}\delta({\bf x}-{\bf x}').\label{npois}
\end{align}
The infinitesimal generator of gauge transformations $G[\chi]$ is defined using the constraints by
\begin{align}\label{gsymm}
G[\chi] = \int d^3 x \, \left[C_0 {\dot \chi} + C_1\chi\right]
\end{align}
in which $\chi$ is arbitrary. Specifically, a gauge transformation of the potentials is given by ${\bf A}+\{G[\chi],{\bf A}\}={\bf A}+\nabla \chi$ and $A_0+\{G[\chi],A_0\}=A_0-{\dot \chi}$. The naive Hamiltonian is the sum of material-kinetic and electromagnetic energies, plus the generator of gauge transformations,
\begin{align}\label{H}
H=& {\dot {\bf r}}\cdot {\bf p}  + \int d^3 x \, [{\dot A}_0({\bf x}) {\tilde \Pi}_0 ({\bf x}) + {\dot {\bf A}}({\bf x})\cdot {\tilde {\bm \Pi}}({\bf x})] - L\nonumber \\ =& {1\over 2}m{\dot {\bf r}}^2 +{1\over 2}\int d^3 x \,  \left[{\bf E}({\bf x})^2+{\bf B}({\bf x})^2\right]+ G[A_0] \nonumber  \\ 
=& {1\over 2m}\left[{\bf p}-q{\bf A}({\bf r})\right]^2+{1\over 2}\int d^3 x\, \left[{\tilde {\bm \Pi}}({\bf x})^2+{\bf B}({\bf x})^2\right]\nonumber \\  &+ G[A_0],
\end{align}
in which $A_0$ and ${\dot A}_0$ are a Lagrange multipliers for the constraints $C_1$ and $C_0$ respectively. Time evolution of an observable $f$ is determined by the Hamilton equation ${\dot f}=\{f,H\}$. In particular, this gives back the generalised velocities as $m{\dot {\bf r}} = \{m{\bf r},H\}= {\bf p}-q{\bf A}({\bf r})$, ${\dot {\bf A}}=\{{\bf A},H\}= {\tilde {\bm \Pi}}-\nabla A_0$ and ${\dot A}_0=\{A_0,H\}$ consistent with Eqs.~(\ref{pcm})-(\ref{picm}).

The third and final constraint necessary is a gauge-fixing constraint. For this we will take $C_2=0$ as defined in Eq.~(\ref{c2}), which as we remarked in Sec.~\ref{back}, is sufficiently general to accommodate both the Coulomb and Poincar\'e gauges. The reduced phase-space within which the constraints hold is the physical state space of the Hamiltonian theory. On this subspace $G[A_0]=0$ and the Hamiltonian is the total energy. Since $\{C_0,C_1\}=0$ and $\{C_0,C_2\}=0$, the dynamics of both $A_0$ and ${\tilde \Pi}_0$ can be confined entirely to the complement of the physical state space and so the constraint $C_0=0$ can be imposed immediately. This removes $A_0$ and ${\tilde \Pi}_0$ from the formalism completely. The Poisson brackets $C_{ij}({\bf x},{\bf x}'):=\{C_i({\bf x}),C_j({\bf x}')\}$ of the remaining two constraints form a non-singular matrix with inverse
\begin{align}
C^{-1}({\bf x},{\bf x}')=\delta({\bf x}-{\bf x}')\left( {\begin{array}{cc}
 0 & 1 \\
-1 & 0 
 \end{array} } \right).
\end{align}
The equal-time {\em Dirac bracket} is defined by
\begin{align}\label{dbrack}
\{\cdot,\cdot\}_D &:= \nonumber \\ \{\cdot,\cdot\}& - \int d^3 x \int d^3 x' \, \{\cdot ,C_i({\bf x})\}C_{ij}^{-1}({\bf x},{\bf x}')\{C_j({\bf x}'),\cdot\}
\end{align}
where repeated indices are summed. Like the Poisson bracket the Dirac bracket is a Lie bracket, but unlike the Poisson bracket, it will yield the correct equations of motion when used in conjunction with the Hamiltonian, even once the constraints $C_i=0$ have been imposed.

Hereafter we denote contravariant indices with subscripts. The nonzero Dirac brackets between the remaining canonical variables are easily computed to be \cite{woolley_r._g._charged_1999}
\begin{align}
\{r_i,p_j\}_D &=\delta_{ij}\label{rp}, \\  \{A_i({\bf x}),{\tilde \Pi}_j({\bf x}')\}_D &=\delta_{ij}\delta({\bf x} - {\bf x}') + \nabla^{\bf x}_i{\rm g}_j({\bf x}',{\bf x})\label{dbapi}, \\  \{p_i,{\tilde \Pi}_j({\bf x})\}_D &= q \nabla^{\bf r}_i{\rm g}_j ({\bf x},{\bf r})=-\nabla_i^{\bf r} P_j({\bf x}),\label{pPi}
\end{align}
where ${\bf P}$ is defined in Eq.~(\ref{P2}). These Dirac brackets are consistent with those given in Ref.~\cite{rousseau_quantum-optics_2017}.
Quantisation of the theory may now be carried out via the replacement $\{\cdot,\cdot\}_D\to-i[\cdot,\cdot]$. The construction of the quantum theory is complete. However, so far only the Dirac brackets of the fields ${\bf A}$ and ${\tilde {\bm \Pi}}$ have been determined and as operators these fields provide an inconvenient expression of the quantum theory, due to Eq.~(\ref{pPi}). This feature is noted in Ref.~\cite{vukics_gauge-invariant_2021} and its response Ref.~\cite{rousseau_reply_2018}. The ensuing lack of commutativity between ${\bf p}$ and ${\tilde {\bm \Pi}}$ within the final quantum theory, implies that the canonical pairs $({\bf r},{\bf p})$ and $({\bf A},{\tilde {\bm \Pi}})$ do not define separate (``matter" and ``light") quantum subsystems. In order to understand the light-matter quantum state space as a tensor-product of a material Hilbert space $[L^2({\mathbb R}^3)]$ and a photonic Fock space $[{\cal F}(L^2({\mathbb R}^3;{\mathbb C}^2))]$, we must identify material and photonic canonical degrees of freedom that are in involution with respect to the Dirac Bracket.

It is straightforward to construct canonical operator pairs that define quantum subsystems by imposing the constraints. The constraint $C_1:=\nabla \cdot {\tilde {\bm \Pi}} +\rho = 0$ uniquely fixes ${\tilde {\bm \Pi}}_{\rm L}$ as a function of ${\bf r}$ through the charge density $\rho({\bf x})$ given in Eq.~(\ref{rho}) as
\begin{align}
{\tilde {\bm \Pi}}_{\rm L}({\bf x}) = -\int d^3 x' {\bf g}_{\rm L}({\bf x},{\bf x}')\rho({\bf x}')  = {\bf P}_{\rm L}({\bf x}) =-{\bf E}_{\rm L}({\bf x}) 
\end{align}
where ${\bf g}_{\rm L}$ is defined in Eq.~(\ref{gl}). The constraint $C_2=0$ implies that ${\bf A}$ can be written as in Eq.~(\ref{Aarb}) and so it is fully determined by ${\bf A}_{\rm T}$ and ${\bf g}_{\rm T}$. We now define the momentum ${\bm \Pi}$ by
\begin{align}\label{pitilde}
{\bm \Pi} &= {\tilde {\bm \Pi}} - {\bf P} = {\tilde {\bm \Pi}}_{\rm T} - {\bf P}_{\rm T}=-{\bf E} - {\bf P}=-{\bf E}_{\rm T} - {\bf P}_{\rm T},
\end{align}
where the second, third, and fourth equalities hold for $C_1=0$. Since immediately we have that $\{p_i,P_j({\bf x})\}_D = -\nabla_i^{\bf r} P_j({\bf x})$, it follows from Eq.~(\ref{pPi}) that
\begin{align}
\{p_i,\Pi_j({\bf x})\}_D  =\{p_i,{\tilde \Pi}_j({\bf x})\}_D -  \{p_i,P_j({\bf x})\}_D = 0.\label{dbppicgt}
\end{align}
Thus, the only non-zero Dirac Brackets of the variables within the set $\{{\bf r},\,{\bf p},\,{\bf A}_{\rm T},\,{\bm \Pi}\}$ are
\begin{align}
\{r_i,p_j\}_D & = \delta_{ij},\label{rp2}\\
\{A_{{\rm T},i}({\bf x}),\Pi_j({\bf x}')\}_D &=\delta_{ij}^{\rm T}({\bf x} - {\bf x}'),\label{trd}
\end{align}
where the second bracket follows immediately from Eq.~(\ref{dbapi}) and $\{A_i({\bf x}),P_j({\bf x}')\}_D=0$. On the physical space of states the relation between the sets  $\{{\bf r},\,{\bf p},\,{\bf A},\,\tilde{\bm \Pi}\}$ and $\{{\bf r},\,{\bf p},\,{\bf A}_{\rm T},\,{\bm \Pi}\}$ is invertible. ${\bf A}$ is given in terms of ${\bf A}_{\rm T}$ by Eq.~(\ref{Aarb}) while ${\bf A}_{\rm T}$ is given in terms of ${\bf A}$ by projecting onto the transverse part. The equation (\ref{pitilde}) defining $\tilde{\bm \Pi}$ is clearly invertible as $\tilde{\bm \Pi}={\bm \Pi}+{\bf P}$. On the physical state space, the algebraic relations (\ref{rp})-(\ref{pPi}) for the set $\{{\bf r},\,{\bf p},\,{\bf A},\,\tilde{\bm \Pi}\}$ hold if and only if the relations (\ref{rp2}) and (\ref{trd}) hold for the set $\{{\bf r},\,{\bf p},\,{\bf A}_{\rm T},\,{\bm \Pi}\}$. Any given operator, such as the Hamiltonian, can be written in terms of either set, and calculations are then performed using the corresponding algebraic relations.

The set $\{{\bf r},\,{\bf p},\,{\bf A}_{\rm T},\,{\bm \Pi}\}$ define matter and light quantum subsystems upon quantisation. Material operators ${\bf r}$ and ${\bf p}$ act within $L^2({\mathbb R}^3)$. On the composite space, $L^2({\mathbb R}^3)\otimes {\cal F}(L^2({\mathbb R}^3;{\mathbb C}^2))$, they have the form ${\bf r}\otimes I_{\rm ph}$ and ${\bf p}\otimes I_{\rm ph}$ where $I_{\rm ph}$ is the identity on the photonic Fock space ${\cal F}(L^2({\mathbb R}^3;{\mathbb C}^2))$. The operators ${\bf A}_{\rm T}$ and ${\bm \Pi}$ act within ${\cal F}(L^2({\mathbb R}^3;{\mathbb C}^2))$ and on the composite space have the form $I_m\otimes {\bf A}_{\rm T}$ and $I_m\otimes {\bm \Pi}$ where $I_m$ is the identity on $L^2({\mathbb R}^3)$. Photon states which span the photonic Fock space are defined using the photonic operator \cite{cohen-tannoudji_photons_1989}
\begin{align}
a_\lambda({\bf k})= { {\bf e}_\lambda({\bf k})\over \sqrt{2k}}\cdot [k{\bf A}_{\rm T}({\bf k}) + i {\bm \Pi}({\bf k})]
\end{align}
where $\lambda=1,2$ specifies the two orthogonal (polarisation) directions orthogonal to ${\bf k}$, such that $\{{\bf e}_1,{\bf e}_2, {\hat {\bf k}}\}$ is an orthonormal triad of unit vectors. The photonic operators satisfy \cite{cohen-tannoudji_photons_1989}
\begin{align}
[a_\lambda({\bf k}),a^\dagger_{\lambda'}({\bf k}')] = \delta_{\lambda\lambda'}\delta({\bf k}-{\bf k}').
\end{align}
consistent with Eq.~(\ref{trd}).


If we choose the Poincar\'e-gauge, that is, if we let ${\bf g}_{\rm T}({\bf x},{\bf x}')=-\int_0^1 d\lambda \, {\bf x}'\cdot \delta^{\rm T}({\bf x}-\lambda{\bf x}')$ as in Eq.~(\ref{pgt}), then the momentum ${\bm \Pi}$ is equal to $-{\bf E}_{\rm T}-{\bf P}_{\rm T}$ where ${\bf P}_{\rm T}$ is the multipolar transverse polarisation. In this case $-{\bm \Pi}={\bf D}_{\rm T}$ is nothing but the standard multipolar transverse displacement field. The Poincar\'e-gauge Hamiltonian is given by Eq.~(\ref{H}) in which ${\bf A}({\bf x})$ is the Poincar\'e-gauge potential and $-{\tilde {\bm \Pi}}={\bf E}$ is the total electric field. This result coincides with the final result of Ref.~\cite{rousseau_reply_2018} (Eq.~(12)  therein). As in Ref.~\cite{rousseau_reply_2018}, all algebraic relations between objects appearing in the Poincar\'e-gauge Hamiltonian are fully specified by the Dirac Brackets in Eqs.~(\ref{rp})-(\ref{pPi}). The authors of Ref.~\cite{rousseau_reply_2018} remark that when expressed in terms of the transverse vector potential ${\bf A}_{\rm T}$ and the longitudinal part ${\bf A}_{\rm L}=\nabla \chi_g[{\bf A}_{\rm T}]$ [in which ${\bf g}_{\rm T}$ is given by Eq.~(\ref{pgt})], the Poincar\'e-gauge Hamiltonian is not the multipolar Hamiltonian. However, when expressed in terms of the set $\{{\bf r},\,{\bf p},\,{\bf A}_{\rm T},\,{\bm \Pi}\}$ the Poincar\'e-gauge Hamiltonian is given by
\begin{align}\label{Hmult}
H =& {1\over 2m}\left[{\bf p}-q{\bf A}({\bf r})\right]^2+{1\over 2}\int d^3 x \, {\bf P}_{\rm L}({\bf x})^2 \nonumber \\ &+ {1\over 2}\int d^3 x\, \left[[{\bm \Pi}({\bf x})+{\bf P}_{\rm T}({\bf x})]^2+{\bf B}({\bf x})^2\right]
\end{align}
wherein the Poincar\'e-gauge potential would usually be expressed in terms of ${\bf B}$ as
\begin{align}
{\bf A}({\bf r}) = -\int_0^1 d\lambda\, \lambda{\bf r} \times {\bf B}(\lambda {\bf r})
\end{align}
and where the fields ${\bf P}_{\rm T}$ and ${\bm \Pi}=-{\bf D}_{\rm T}$ are the multipolar transverse polarisation and transverse displacement field respectively. The contribution
\begin{align}
&{1\over 2}\int d^3x \, {\bf P}_{\rm L}({\bf x})^2 = {1\over 2}\int d^3x \, {\bf E}_{\rm L}({\bf x})^2  \nonumber \\ &= \int d^3 x\int d^3 x'\, {\rho({\bf x})\rho({\bf x}')\over 8\pi|{\bf x}-{\bf x}'|}= V_{\rm Coul}
\end{align}
is the energy of the longitudinal electric field which by Gauss' law $C_1=0$ is the Coulomb energy of the charge distribution. It includes the  nuclear binding energy $-q^2/(4\pi|{\bf r}|)$ as well as the infinite Coulomb self-energies of the constituent charges. Eq.~(\ref{Hmult}), which is nothing but the Poincar\'e-gauge Hamiltonian expressed in terms of a convenient set of variables and on the physical subspace, is identical to the well-known multipolar Hamiltonian of textbook nonrelativistic QED \cite{cohen-tannoudji_photons_1989,craig_molecular_1998}. 
Refs.~\cite{rousseau_quantum-optics_2017,rousseau_reply_2018} conclude that when written in terms of ${\bf A}_{\rm T}$ and ${\tilde {\bm \Pi}}=-{\bf E}_{\rm T}-{\bf E}_{\rm L}$, the Poincar\'e-gauge Hamiltonian is not the multipolar Hamiltonian, because ${\tilde {\bm \Pi}}_{\rm T}$ equals $-{\bf E}_{\rm T}$ rather than $-{\bf D}_{\rm T}$ and so the 
momentum ${\tilde {\bm \Pi}}_{\rm T}$ is {\em not} the well-known canonical momentum encountered in textbook multipolar theory. However, 
what is required in order that the two theories coincide is that ${\bm \Pi}_{\rm T}=-{\bf D}_{\rm T}$, and this is the case. Indeed, as we have shown, this equality is {\em implied} by the equality ${\tilde {\bm \Pi}}_{\rm T}=-{\bf E}_{\rm T}$, which therefore proves that the two theories are identical rather than disparate. 

We have shown that when the constraints are satisfied $C_1=0=C_2$, the Poincar\'e-gauge Hamiltonian and the multipolar Hamiltonian are one and the same. This Hamiltonian results from the Poincar\'e gauge a.k.a multipolar-gauge choice of ${\bf g}_{\rm T}$ given in Eq.~(\ref{pgt}), which specifies that ${\bf x
}\cdot {\bf A}({\bf x})=0$. The route we have taken has been to derive the Hamiltonian on the extended space, constructing the algebra of operators using Dirac's method, and to then reduce the theory to the physical subspace defined by $C_1=0=C_2$. Alternatively one can use the constraints at the outset to show that the Poincar\'e gauge and multipolar Lagrangians are one and the same. Specifically, letting ${\bf A}={\bf A}_{\rm T}+\nabla \chi$ and $A_0=\phi-{\dot \chi}$ where $\chi({\bf x}) = -\int_0^1 d\lambda \,{\bf x}\cdot {\bf A}_{\rm T}({\bf x})$ defines the Poincar\'e gauge, and substituting these expressions into Eq.~(\ref{L}) yields the Poincar\'e gauge Lagrangian
\begin{align}
L_p =& {1\over 2}m{\dot {\bf r}}^2 -\int d^3 x\, j^\mu A_\mu - {1\over 4}\int d^3 x F_{\mu\nu}F^{\mu\nu} \nonumber \\
=& {1\over 2}m{\dot {\bf r}}^2-V_{\rm Coul}+\int d^3 x \, {\bf J}\cdot {\bf A}_{\rm T} \nonumber \\ &+ {1\over 2}\int d^3 x\, [{\bf E}_{\rm T}^2-{\bf B}^2] + \int d^3 x\, [{\bf J}\cdot \nabla\chi +\rho {\dot \chi}]\nonumber \\
=& L_c + {d\over dt}\int d^3x\, \rho \chi
\end{align}
where $L_c$ is the Coulomb-gauge Lagrangian and we have used ${\dot \rho}=-\nabla\cdot {\bf J}$. We have also used $C_1=0$ to give ${\bf E}_{\rm L}=\nabla(\nabla^{-2}\rho)$. Moreover, we have that
\begin{align}
{d\over dt} \int d^3 x \rho \chi = -{d\over dt}\int d^3x\, {\bf P}\cdot {\bf A}_{\rm T} = -{d\over dt}\int d^3x\, {\bf P}_{\rm T}\cdot {\bf A}_{\rm T}
\end{align}
where ${\bf P}$ is the multipolar polarisation given in Eq.~(\ref{multP}). Thus, $L_p =L_c - {d\over dt}\int d^3 x {\bf P}_{\rm T}\cdot {\bf A}_{\rm T} =: L_{\rm mult}$ where $L_{\rm mult}$ is the standard textbook multipolar Lagrangian found as a Lagrangian Power-Zienau-Woolley (PZW) transformation of $L_c$ \cite{craig_molecular_1998,cohen-tannoudji_photons_1989}. Taking ${\bf A}_{\rm T}$ as coordinate, the multipolar Hamiltonian can now be found from $L_p=L_{\rm mult}$ via Legendre transformation as is the route usually adopted in textbooks \cite{craig_molecular_1998,cohen-tannoudji_photons_1989}.

\section{Discussion and generalisation}\label{discuss}

The misunderstanding stems from a one-to-two usage of the name ``canonical momentum". {\em Apparent} disagreement between results occurs because different authors use this label for different fields. In multipolar QED we call ${\bm \Pi}=-{\bf D}_{\rm T}$ the canonical momentum, because in the final unconstrained theory it is conjugate to ${\bf A}_{\rm T}$ [in the sense of Eq.~(\ref{trd})] and it commutes with ${\bf r}$ and ${\bf p}$. On the other hand, when we follow Dirac's method of quantisation (as above and as in Refs.~\cite{rousseau_quantum-optics_2017,rousseau_reply_2018}) the object termed ``canonical momentum" is ${\tilde {\bm \Pi}}=-{\bf E}$, because in the starting naive (constrained) theory this momentum is conjugate to ${\bf A}$ [in the sense of Eq.~(\ref{npois})] and it commutes with ${\bf r}$ and ${\bf p}$. Thus, the same name ``canonical momentum" has been used for distinct fields that are not equal but that are instead related by Eq.~(\ref{pitilde}). Both of these nomenclatures are reasonable, but adopting them both simultaneously will inevitably cause confusion. We must recognise that neither ${\tilde {\bm \Pi}}$ nor ${\tilde {\bm \Pi}}_{\rm T}$ equals ${\bm \Pi}$ in general. This fact does not imply that Poincar\'e-gauge and multipolar QED are not the same, and in fact, by taking into account the relationship between ${\tilde {\bm \Pi}}$ and ${\bm \Pi}$ one can prove that the two theories are identical, as we have done.

More generally, the Hamiltonian in Eq.~(\ref{H}) expressed on the physical subspace in terms of the set $\{{\bf r},\,{\bf p},\,{\bf A}_{\rm T},\,{\bm \Pi}\}$ is found from Eqs.~(\ref{H}) and (\ref{pitilde}) to be
\begin{align}\label{Hg}
H[{\bf g}_{\rm T}] =& {1\over 2m}\left[{\bf p}-q{\bf A}_g({\bf r})\right]^2 \nonumber \\ &+{1\over 2}\int d^3 x\, \left[[{\bm \Pi}({\bf x})+{\bf P}_g({\bf x})]^2+{\bf B}({\bf x})^2\right]
\end{align}
where ${\bf A}_g$ and ${\bf P}_g$ are the vector potential and polarisation on the physical subspace given by Eqs.~(\ref{Aarb}) and (\ref{P2}) respectively as
\begin{align}
&{\bf A}_g({\bf x})={\bf A}_{\rm T}({\bf x})+\nabla\int d^3 x' {\bf g}({\bf x'},{\bf x})\cdot {\bf A}_{\rm T}({\bf x}'),\\
&{\bf P}_g({\bf x}) = -\int d^3 x' {\bf g}({\bf x},{\bf x}')\rho({\bf x}').
\end{align}
The gauge is fully determined via a choice of ${\bf g}_{\rm T}$, which uniquely determines ${\bf g}$. Upon quantisation the Hamiltonians of different gauges ${\bf g}_{\rm T}$ and ${\bf g}_{\rm T}'$ are unitarily related as
\begin{align}\label{hamrel}
H[{\bf g}_{\rm T}'] = U_{gg'}H[{\bf g}_{\rm T}] U_{gg'}^\dagger
\end{align}
where the {\em gauge-fixing} transformation $U_{gg'}$ is defined over the physical Hilbert space by
\begin{align}\label{ugg}
U_{gg'}&:= \exp\left(-i\int {\rm d}^3x \, \big[\chi_g({\bf x})-\chi_{g'}({\bf x})\big]\rho({\bf x}) \right),\\
&= \exp\left(i\int {\rm d}^3x \, \big[{\bf P}_g({\bf x})-{\bf P}_{g'}({\bf x})\big]\cdot {\bf A}_{\rm T}({\bf x}) \right).
\end{align} 
Equation~(\ref{hamrel}) can be obtained immediately from
\begin{align}
&U_{gg'}[{\bf p}-q{\bf A}_g({\bf r})]U_{gg'}^\dagger = {\bf p}-q{\bf A}_{g'}({\bf r}),\\
&U_{gg'}[{\bm \Pi}+{\bf P}_{g{\rm T}}]U_{gg'}^\dagger = {\bm \Pi}+{\bf P}_{g'{\rm T}},
\end{align}
showing that $U_{gg'}$ implements the gauge-change ${\bf g}_{\rm T}\to{\bf g}_{\rm T}'$ within the Hamiltonian via transformation of the momenta ${\bf p}$ and ${\bm \Pi}$. The Coulomb and Poincar\'e (multipolar)-gauge Hamiltonians are special cases obtained by making the Coulomb and Poincar\'e-gauge choices of ${\bf g}_{\rm T}$ respectively. The unitary gauge-fixing transformation between these particular gauges is called the Power-Zienau-Woolley (PZW) transformation. It is given according to Eq.~(\ref{ugg}) by
\begin{align}
U = \exp\left(-i\int {\rm d}^3x \, {\bf P}({\bf x})\cdot {\bf A}_{\rm T}({\bf x}) \right),
\end{align} 
where ${\bf P}$ is the multipolar polarisation. This is the textbook expression for the PZW transformation \cite{cohen-tannoudji_photons_1989,craig_molecular_1998}. 

We remark that previous authors have concluded that the PZW transformation is not a gauge-transformation \cite{andrews_perspective_2018}, but this conclusion is not at odds with our findings. The PZW transformation is not a {\em gauge-symmetry} transformation, which is defined as a transformation that acts directly on the potential $A$ to implement a gauge transformation $A\to A-d\chi$ within the starting constrained theory. 
The PZW transformation {\em is} however, an example of a  {\em gauge-fixing} transformation $U_{gg'}$, which is defined as a unitary transformation acting within the final unconstrained theory to transform from one fixed-gauge realisation of the physical state space to a different fixed-gauge realisation. There is clearly a distinction between gauge-symmetry and gauge-fixing transformations in Hamiltonian QED, and the distinction provides unambiguous clarification of the relation between the PZW transformation and gauge-freedom.


\section{Conclusions}\label{conc}

Refs.~\cite{rousseau_quantum-optics_2017,rousseau_reply_2018} express the Poincar\'e-gauge theory in terms of the Poincar\'e-gauge potential ${\bf A}$ and the momentum ${\tilde {\bm \Pi}}$ (see for example Eq.~(12) of Ref.~\cite{rousseau_reply_2018}). The multipolar framework is the same theory expressed in terms of different fields ${\bf A}_{\rm T}$ and ${\bm \Pi}$, which are more convenient for use within the quantum theory. We have verified this via the explicit construction of Dirac Brackets as are also derived in Refs.~\cite{rousseau_quantum-optics_2017,rousseau_reply_2018}. Before now such a demonstration had not been clearly provided within the literature. Indeed, as well as being unrecognised in Refs.~\cite{rousseau_quantum-optics_2017,rousseau_reply_2018}, the distinction between ${\tilde {\bm \Pi}}$ and ${\bm \Pi}$ is perhaps also obfuscated elsewhere. For example, the constraint $C_2=0$ used in this article has been employed by Woolley in Ref.~\cite{woolley_r._g._charged_1999}, which summarises the approach developed much earlier by the same author in Refs.~\cite{woolley_molecular_1971,woolley_reformulation_1974,woolley_non-relativistic_1975}. Woolley constructs the Dirac brackets for the theory, but typically adopts the notation ${\bf E}^\perp$ for $-{\bm \Pi}$. We emphasise that $-{\bm \Pi}$ does not represent the transverse electric field except when ${\bf g}_{\rm T}={\bf 0}$ (Coulomb-gauge). We emphasise that the distinction between Coulomb-gauge and multipolar QED is no more or less than a distinction between gauge choices. As we have shown, all gauges of this type are related by unitary gauge-fixing transformations and they are all equivalent to each other. Multipolar QED in particular, is strictly equivalent to Coulomb-gauge QED and is identical to the Poincar\'e-gauge theory.

More generally, we have shown that the conventional Hamiltonians of nonrelativistic QED used in light-matter physics and quantum optics, are fixed-gauge cases of the general Hamiltonian $H$ in Eq.~(\ref{H}) once it has been restricted to the physical subspace and written in terms of convenient canonical variables that possess the algebraic properties in Eqs.~(\ref{rp2}) and (\ref{trd}). This writing of $H$ is nothing but $H[{\bf g}_{\rm T}]$ given in Eq.~(\ref{Hg}). All instances of this Hamiltonian are unitarily equivalent, being related by unitary transformations confined to the physical subspace. They are therefore all physically equivalent. This is nothing but canonical QED's expression of {\em gauge-invariance}.

The Hamiltonian $H[{\bf g}_{\rm T}]$ is consistent with all of the required physical constraints and yields all of the required dynamical equations of motion for {\em any} choice of gauge ${\bf g}_{\rm T}$. This fact alone justifies the use of $H[{\bf g}_{\rm T}]$ for any choice of ${\bf g}_{\rm T}$ and in particular it justifies the use of the multipolar theory. While it is satisfying to connect a Hamiltonian to a Lagrangian description, there is no further justification for the latter than that it produces the correct equations of motion. Thus, $H[{\bf g}_{\rm T}]$ is no less fundamental than the Lagrangian $L$ from which it happened to be derived. 
In this sense, any criticism of Hamiltonian multipolar QED as less physical than another description is immediately unfounded because whether or not one can derive it from some other description, one can certainly verify that it produces the required equations of motion. The pertinent question therefore, is whether or not the Hamiltonians commonly used in practice, all of which produce the correct equations of motion, are physically equivalent in the general sense. Fortunately, one can show (as we have) that they are, because the quantum-theoretic definition of physical equivalence is unitary equivalence. In particular, our results rigorously solidify the multipolar theory and we have clarified how alternative non-relativistic gauges are related to the Coulomb-gauge through unitary gauge-fixing transformations expressed in terms of the gauge-invariant transverse vector-potential.\\

\noindent {\em Acknowledgement}. This work was supported by the UK Engineering and Physical Sciences Research Council, Grants No. EP/N008154/1 and No. EP/V048562/1.

\bibliography{lib.bib}

\end{document}